\documentclass[amsmath,amssymb,twocolumn,aps,prl,showpacs,superscriptaddress]{revtex4-1}
\usepackage{graphicx}
\usepackage{dcolumn}

\newcommand{\degree}{\ensuremath{^{\circ}}\,\,}

\newcommand{\LCO}{\ensuremath{\text{La}_{2}\text{CuO}_4}\,\,}
\newcommand{\SCOC}{\ensuremath{\text{Sr}_{2}\text{CuO}_{2}\text{Cl}_2}\,\,}

\newcommand{\YBCO}{\ensuremath{\text{YBa}_{2}\text{Cu}_{3}\text{O}_{6}}\,\,}
\newcommand{\CuIon}{\ensuremath{\text{Cu}^{\text{2+}}\,}}

\newcommand{\etal}{\emph{et al.\,\,}}

\begin{document}
\title{High-Energy Continuum of Magnetic Excitations in the Two-Dimensional 
    Quantum Antiferromagnet \SCOC}

\author{K. W. Plumb}
\affiliation{Department of Physics, University of Toronto, Toronto, Ontario,
Canada M5S 1A7}

\author{A.~T.~Savici}
\affiliation{Neutron Data Analysis and Visualization Division, Oak Ridge
National Laboratory, Oak Ridge, Tennessee 37831, USA}

\author{G.~E.~Granroth}
\affiliation{Neutron Data Analysis and Visualization Division, Oak Ridge
National Laboratory, Oak Ridge, Tennessee 37831, USA}
\affiliation{Quantum Condensed Matter Division, Oak Ridge National Laboratory,
Oak Ridge, Tennessee 37831, USA}

\author{F.C. Chou}
\affiliation{Center for Condensed Matter Sciences, National Taiwan University,
Taipei, 10617 Taiwan}

\author{Young-June Kim}
\affiliation{Department of Physics, University of Toronto, Toronto, Ontario,
 Canada M5S 1A7}

\date{\today}

\begin{abstract}
We have measured the magnetic excitation spectrum of the model square-lattice
spin-1/2 antiferromagnet \SCOC over a broad range of energy and momentum using
high-resolution inelastic neutron scattering (INS). The magnon dispersion along
the zone boundary was accurately measured to be a 43~meV between $(1/2,0)$ and
$(3/4,1/4)$ indicating the importance of coupling beyond nearest neighbors in
the spin Hamiltonian. We observe a strong momentum dependent damping of the
zone-boundary magnons at $(1/2,0)$ revealing a high energy continuum of
magnetic excitations. A direct comparison between our INS measurements and
resonant inelastic X-ray scattering (RIXS) measurements shows that the RIXS
spectrum contains significant contributions from higher energy excitations not
previously considered. Our observations demonstrate that this high-energy
continuum of magnetic fluctuations is a ubiquitous feature of insulating
monolayer cuprates, apparent in both inelastic neutron and light scattering
measurements.
\end{abstract}


\maketitle
In the continued search for new states of matter, characterized by 
unconventional collective modes, model systems play an essential role advancing 
quantitative understanding of  novel theoretical ideas and state of the art 
experimental techniques. The value of a well-established model system is that it 
allows researchers to focus on solving, rather than defining, the problem. The 
spin-1/2 Heisenberg model on a square lattice is one of the most extensively 
studied systems of quantum magnetism.  Interest in this model has been largely 
stimulated by its relevance to the parent compounds of the high temperature 
cuprate superconductors~\cite{Lee:06}, \LCO is perhaps the best known example in 
this regard~\cite{Kastner:Rev:1998}.  While the long-wavelength spin excitations 
in \LCO are correctly described by a renormalized spin wave theory within the 2D 
Heisenberg model, experiments probing the high energy spin fluctuations have 
shown that the nearest-neighbor Heisenberg model cannot capture details of the 
spin excitation spectrum at short wavelengths. For example, the broad 
asymmetrical line-shape of the two-magnon Raman scattering cannot be explained 
by calculations based on spin-wave theory~\cite{Sugai:90,Katanin:03}. More 
recently, inelastic neutron scattering (INS) measurements conducted by Headings 
\etal reveal a high energy momentum dependent continuum of magnetic excitations 
in \LCO~\cite{Headings:2010}.

In this letter, we report high resolution INS measurements of the magnetic 
excitation spectrum of another well-known 2D model system, the insulating 
cuprate \SCOC~\cite{Greven:1994,Greven:95,Thurber:97}. We observe a strong 
momentum dependent damping of the zone boundary magnons indicating the presence 
of a high-energy continuum of magnetic excitations in \SCOC\!\!. Furthermore, we 
directly compare the INS spectra with previously reported resonant inelastic 
X-ray scattering (RIXS) results~\cite{Guarise:2010}.  Importantly, the RIXS and 
INS measurements are in agreement over a large portion of the Brillouin zone 
\emph{except} in the neighborhood of $(1/2,0)$ where RIXS overestimates the zone 
boundary energy by $\sim$25~meV. This deviation occurs in precisely the region 
of phase-space where the continuum scattering carries significant spectral 
weight. The strong coupling of the RIXS cross-section to the momentum dependent 
continuum, in conjunction with the broad energy resolution of RIXS, accounts for 
discrepancies between the zone-boundary magnon dispersion measured by RIXS and 
INS\@.  Thus, the high-energy momentum dependent continuum of spin fluctuations 
previously observed in \LCO~\cite{Headings:2010} appears to be a general feature 
of the magnetic excitation spectrum in the parent cuprates, evident in both 
inelastic neutron and light scattering measurements.

For our measurements, large single crystals of \SCOC were grown from flux 
following identical synthesis procedures to previous reports~\cite{Miller:90}.  
\SCOC has a tetragonal unit cell, space group I4/mmm, down to at least 5~K with 
lattice parameters, a = b = 3.96 \AA\ and c = 15.53 \AA\, and we index momentum 
transfer $\mathbf{Q} = h\mathbf{a}^{\star} + k\mathbf{b}^{\star} 
+l\mathbf{c}^{\star}$ using the reciprocal lattice of the structural unit cell.  
Measurements were performed on an array of seven co-aligned single crystals with 
a total mass of 17~g, and a mosaic spread of 2\degree\!. The sample was aligned 
with the $\mathbf{c}^{\star}$-axis along the incident neutron wavevector, $k_i$ 
for the duration of the measurement.

INS measurements were performed on the fine resolution Fermi chopper 
spectrometer SEQUOIA, at the Spallation Neutron Source (SNS) at Oak Ridge 
National Laboratory. Measurements of the zone boundary magnetic excitations were 
carried out in two configurations: with Fermi chopper 1 rotating at a frequency 
of 480~Hz phased for an incident energy of 450~meV, and with Fermi chopper 1 at 
300~Hz phased for an incident energy of 150~meV. A T$0$ chopper rotating at 
120~Hz was used to eliminate the fast neutron background 
\cite{Granroth:2006,Granroth:2010}. INS intensities were put on an absolute 
scale by normalization with incoherent elastic scattering from the sample.

\begin{figure}[htb] 
    \includegraphics[]{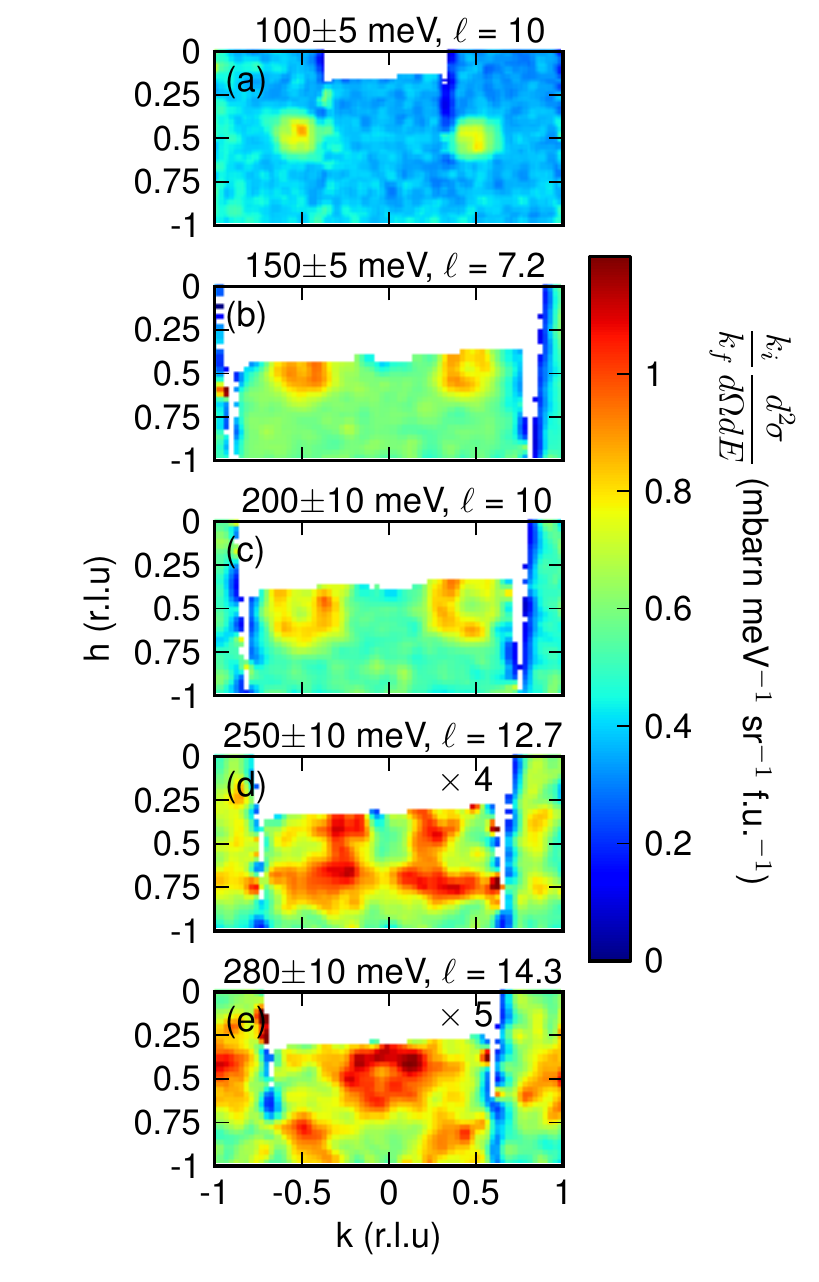}
    \caption{\label{fig:const_E_slices} Representative constant energy transfer
        slices displaying the raw INS intensity at T = 5~K. Data in panel (a)
        was collected with $E_i = 150$~meV and a total proton charge of
        $47.5$~C.  Data in panels (b)-(e) was taken with $E_i = 450$~meV and a
        proton charge of $186$~C.  Data were scaled to absolute units by
        normalizing with incoherent elastic scattering from the sample.}
\end{figure}

Figure~\ref{fig:const_E_slices} shows a set of typical constant energy transfer
slices in the $(h,k)$ plane of the raw neutron scattering intensity. The INS
cross-section is directly proportional to the dynamical structure factor and
high intensity regions are produced when the spectrometer resolution volume
intersects with the spin excitation dispersion surface. In
Fig.~\ref{fig:const_E_slices} cones of spin waves can be seen emerging from
the antiferromagnetic zone centers at $(1/2,1/2)$ which disperse
isotropically at low energies. As the zone-boundary energy is approached around
280 meV [Fig.~\ref{fig:const_E_slices} (e)], the high intensity regions become
localized near the $(1/2,0)$ positions in reciprocal space.

\begin{figure}[htb]
    \includegraphics[]{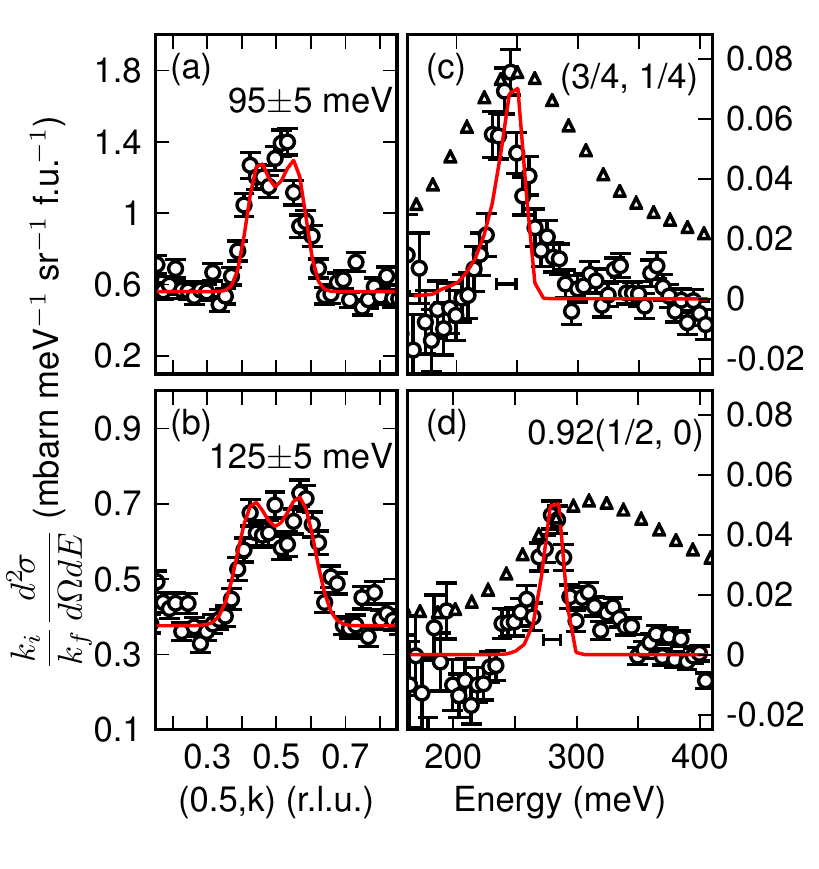} 
    \caption{\label{fig:ZB_cuts} (a)-(b) Constant
        energy cuts near the antiferromagnetic zone center for an incident
        neutron energy of 150 meV. (c)-(d) Constant momentum transfer cuts
        through the zone boundaries integrated over $\pm$0.16 \AA$^{-1}$ for an
        incident neutron energy of 450 meV. An energy dependent background,
        measured near \textbf{Q} = (1,0) has been subtracted from the constant
        momentum transfer cuts. Black solid lines show the time-of- flight
        spectrometer resolution FWHM at the zone boundaries. Red lines are fits
        to the resolution convolved model cross- section, including the
        anisotropic \CuIon$d_{x^2-y^2}$ form-factor \cite{Shamoto:93}.
        Triangles in (c)-(d) are the corresponding RIXS data from reference
        \cite{Guarise:2010}.} 
\end{figure}

\begin{figure}[htb]
    \includegraphics[]{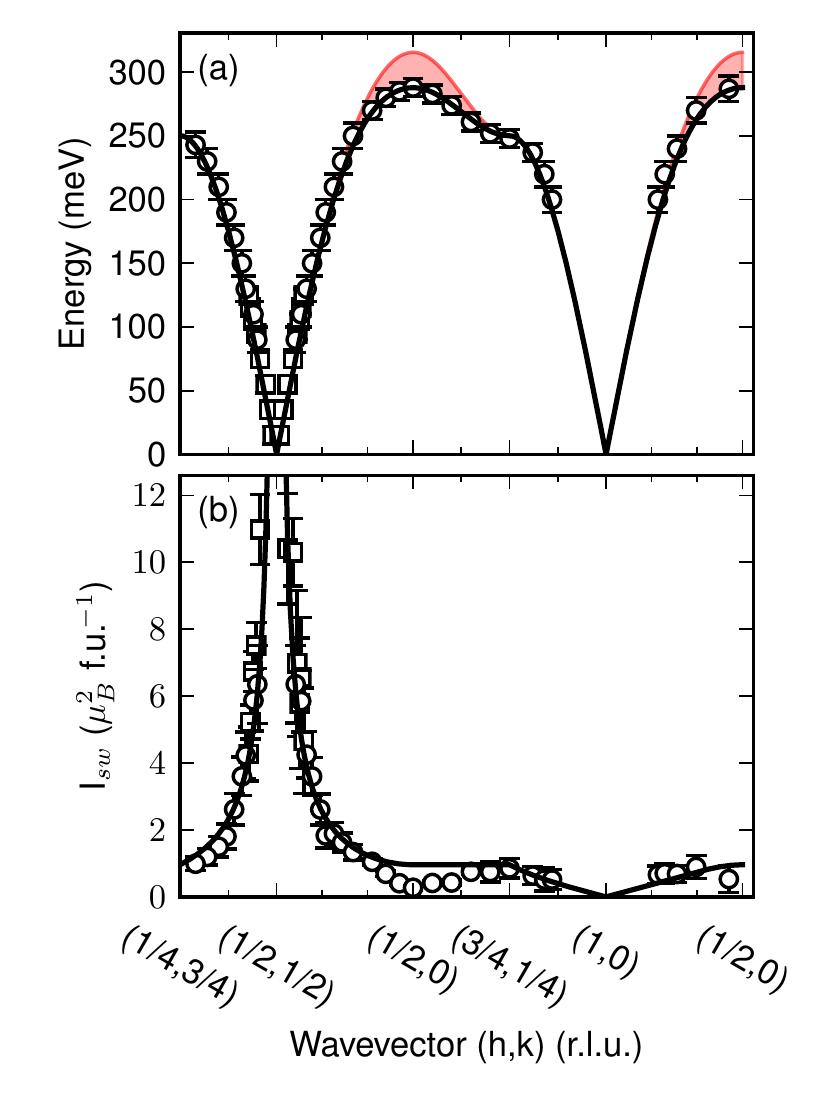}
    \caption{\label{fig:dispersion} (a) Magnon dispersion along high symmetry
        directions measured by INS in \SCOC at T = 5~K. Each point was
        extracted by fitting to individual constant momentum or energy transfer
        cuts. Square points were obtained from the $E_i = 150$~meV data and the
        circles from the $E_i = 450$~meV data. The solid black line is the best
        fit to the spin wave dispersion relation and the red filled area is the
        difference between the RIXS results of \cite{Guarise:2010} and INS
        measurement.  (b) Wave-vector dependence of the integrated spin wave
        intensity. The black solid line is the prediction of linear spin wave
        theory including an intensity renormalization factor $Z_d = 0.48(5)$
        \cite{Lorenzana:2005}.}
\end{figure}

More details of the spin excitations can be extracted from constant energy
transfer and constant momentum transfer cuts through the dispersion surface.
Representative cuts are presented in figure~\ref{fig:ZB_cuts}. The spin wave
dispersion and momentum dependence of intensity were accurately determined by
independently fitting a large number of cuts, spanning the full Brillouin zone,
with a model cross-section based on the Hamiltonian in equation
\eqref{eq:Hamiltonian} convolved with the spectrometer resolution
function~\cite{Tobyfit}.  Figures~\ref{fig:dispersion} (a) and (b) show the
dispersion and spin wave intensities respectively along high symmetry
directions obtained from the cuts.  

The data at small momentum transfers, close to the magnetic zone center, and
over a large portion of the zone boundary are consistent with results from
previous low-energy INS measurements \cite{Greven:95} as well as the recent
RIXS results. We also observe a large, 43~meV, dispersion along the
zone-boundary, between $(1/2,0)$ and $(3/4,1/4)$; however, this is
significantly smaller than the 70~meV zone boundary dispersion reported by a
RIXS study on the identical compound \cite{Guarise:2010}, but still much larger
than the 22~meV dispersion reported for \LCO~\cite{Coldea:2001}.

The observed dispersion can be analyzed following previous work on \LCO which
described the excitation spectrum using classical linear spin wave theory and a
Heisenberg Hamiltonian including higher order couplings in the spin operator
\cite{Coldea:2001,Headings:2010}:
\begin{align}
\label{eq:Hamiltonian}
H = &J\sum_{\langle i,j\rangle}\mathbf{S}_i\cdot\mathbf{S}_j + J^{\prime}\sum_{\langle i,i^{\prime}
\rangle}\mathbf{S}_i\cdot\mathbf{S}_{i^{\prime}} + J^{\prime \prime}\sum_{\langle i,i^{\prime \prime}\rangle}
\mathbf{S}_i\cdot\mathbf{S}_{i^{\prime \prime}} \\*\nonumber
&+ J_{c}\sum_{\langle i,j,k,l \rangle} \lbrace\left(\mathbf{S}_i\cdot\mathbf{S}_j\right)\left(\mathbf{S}_k\cdot\mathbf{S}_l\right)+\left(\mathbf{S}_i\cdot\mathbf{S
}_l\right)\left(\mathbf{S}_k\cdot\mathbf{S}_j\right)\\*\nonumber
&-\left(\mathbf{S}_i\cdot\mathbf{S}_k\right)\left(\mathbf{S}_j\cdot\mathbf{S}_l\right)\rbrace,
\end{align}
where $J$, $J^{\prime}$, and $J^{\prime\prime}$ are the first, second and third
neighbor exchange, and $J_c$ is the ring exchange. Expanding the exchange
constants to fourth order in terms of an effective Coulomb repulsion $U$ and
nearest neighbor hopping $t$, the exchange constants can be determined in terms
of $t$ and $U$ \cite{Coldea:2001,Delannoy:2009}. Fitting $t$ and $U$ within
linear spin wave theory we obtain $t\!=\!0.261(3)$, $U\!=\!1.9(5)$ and $J\!=\!
4t^2/U - 24t^4/U^3\!=\!126(3)$, $J^{\prime}\!=\!J^{\prime
    \prime}\!=\!4t^4/U^3\!=\!2.7(5)$, $J_c\!=\! 80t^4/U^3\!=\!54(3)$. This fit
is shown as a solid black line in Fig.~\ref{fig:dispersion} (a). The momentum
dependence of the magnon spectral weight also predicted by spin wave theory,
including two transverse modes, is compared with the integrated intensity of
the one-magnon signal extracted from INS data in Fig.~\ref{fig:dispersion}
(b). Including an intensity renormalization factor of $Z_d = 0.48(5)$
\cite{Lorenzana:2005}, there is good agreement with the linear spin wave theory
prediction across most of the Brillouin zone except for momentum transfers in
near $(1/2,0)$ where the measured intensity is strongly suppressed compared
with the spin wave theory prediction. The origin of this discrepancy is
revealed by the constant momentum transfer cuts which show a high energy tail
of scattering forming a momentum dependent continuum around $(1/2,0)$. Magnons
at $(1/2,0)$ couple strongly to these higher energy excitations and are thus
heavily damped, reducing the spectral weight contained in the one-magnon peak.
The momentum and energy integrated intensity of the inelastic magnetic
scattering is $\langle M^2_{inelastic} \rangle = 0.95 \pm 0.30\,\,\mu_B^2$
corresponding to approximately $65 \%$ of the inelastic spectral weight
predicted by renormalized spin-wave theory \cite{Lorenzana:2005}.

An identical suppression of the one-magnon intensity around $(1/2,0)$ and 
momentum dependent continuum scattering has previously been observed in both 
\LCO~\cite{Headings:2010} and the square lattice antiferromagnet CFTD 
\cite{Christensen:2007}.  The origin of these high energy magnetic excitations 
is not clear. The broad peak shapes and momentum dependent damping cannot be 
accounted for by including two-magnon process in the INS 
cross-section~\cite{Christensen:2007,Headings:2010}. One possible explanation of 
the high energy spectral weight is the existence of spinons with a d-wave 
dispersion, as proposed in some theoretical models of the 
cuprates~\cite{Ho:2001}. Such  high-energy excitations should also be visible in 
RIXS spectra, which has been shown to couple strongly with the two-and 
four-spinon continuum present in one-dimensional magnetic 
systems~\cite{Glawion:2011, Schlappa:2012}.

RIXS has recently emerged as a powerful probe of collective excitations in
quantum condensed matter systems~\cite{Ament:2011}. RIXS has been used to
measure the dispersion of charge excitons, orbitons, bimagnons, as well as
magnetic excitations in copper oxide based model 
systems~\cite{Hill:2008,Schlappa:2009,Schlappa:2012,Glawion:2011}. The 
observation of momentum resolved collective magnetic excitations has generated a 
great deal of excitement in the research community because RIXS overcomes many 
challenges of INS\@. In particular, measurements can be performed on small 
samples, including thin-films~\cite{Dean:2012}, and on samples containing highly 
absorbing elements such as Ir~\cite{Kim:2012}.  Unfortunately, coarse energy 
resolution currently limits the applicability of this technique to only a 
handful of systems with large magnetic energy scales. RIXS also couples strongly 
to other types of excitations in addition to magnetic collective 
modes~\cite{Ament:2011} which, in conjunction with the coarse energy resolution, 
can severely limit interpretation of experimental data. Since the cross-section 
of INS is well understood, a detailed comparison of magnetic excitation spectrum 
measured by RIXS and INS throughout the full Brillouin zone would be of great 
utility in the development of our understanding of the RIXS cross-section. An 
earlier comparison for \LCO found that the magnon spectrum of \LCO extracted 
from RIXS agrees with neutron scattering data~\cite{Coldea:2001} over most of 
the Brillouin zone~\cite{Braicovich:2009,Braicovich:2010}.  However, this 
comparison was limited in momentum space coverage and did not include the zone 
boundary. 

In figure~\ref{fig:ZB_cuts} (c) and (d) the corresponding RIXS spectra from
reference~\cite{Guarise:2010} are superimposed with the INS data. While the
position of the RIXS and INS peak intensity is in excellent agreement near
$(3/4,1/4)$, there is a significant $\sim$25~meV discrepancy near $(1/2,0)$,
[Fig.~\ref{fig:ZB_cuts} (d)]. The momentum dependence of this discrepancy is
highlighted by the red shaded area in Fig.~\ref{fig:dispersion} (a) which shows
the difference between the magnon dispersion predicted using parameters
extracted from the RIXS data in~\cite{Guarise:2010} and the INS result.

In order to extract the spin wave dispersion, RIXS spectra are typically
analyzed including contributions from one-and two-magnon scattering. Scattering
by the creation of a single magnon is considered to be the dominant process and
the energy position of the spectra maximum corresponds closely to the magnon
energy~\cite{Braicovich:2010,Guarise:2010}. Upon comparison with the INS
lineshape, it is clear that the simple interpretation of the RIXS spectrum in
terms of one-and two-magnon excitations alone is not complete and that there must
be a significant higher energy contribution to the RIXS magnetic spectral weight
near (1/2,0). This was not observed in previous comparisons of RIXS and INS
measurements of the magnon dispersion in \LCO because the RIXS study only
extended up to $\mathbf{q} \approx 0.8(1/2,0)$~\cite{Braicovich:2010}, just
outside the region where we observe the discrepancy between RIXS and INS.

The results presented here show that the momentum dependent continuum 
scattering at $(1/2,0)$~\cite{Headings:2010} is a general feature of the
monolayer parent cuprate compounds and that the magnetic excitation spectrum
cannot be fully described by a renormalized spin wave theory from an ordered
antiferromagnet. This conclusion is further supported by optical measurements
where the deficiency of spin wave theory for describing the short wavelength
magnetic excitation spectra is particularly apparent. Specifically, spin wave
theory fails to explain the asymmetrical line-shape of the two-magnon Raman
scattering~\cite{Sugai:90,Katanin:03} and high energy spectral weight extending
between 400 and 750 meV observed in mid-infrared (midIR) optical absorption 
measurements~\cite{Perkins:98}. MidIR optical absorption spectra of the cuprates 
show a sharp resonance peak near 400 meV~\cite{Perkins:98} which has been 
identified as resulting from a bimagnon plus phonon 
excitation~\cite{Lorenzana:95}; however, the significant spectral weight 
extending from the 400 meV resonance up to 750 meV cannot be explained in this 
context.  Similarly, through careful analysis of the optical conductivity 
line-shape in \YBCO Gr\"{u}ninger \etal conclude that, while the main resonance 
peak observed near 400 meV is well described by the bimagnon plus phonon 
process, the significant spectral weight extending to 750 meV cannot be 
accounted for within spin wave theory~\cite{Gruninger:00}. The inclusion of ring 
exchange and phonon damping terms has also failed to account for the asymmetric 
Raman scattering line shape~\cite{Katanin:03}. A complete description of 
magnetic interactions in the parent cuprates must account for this high energy, 
momentum dependent, continuum of magnetic excitations. 

In conclusion, we have measured the magnetic excitation spectrum of the model
square-lattice spin-1/2 antiferromagnet \SCOC throughout the full Brillouin
zone. Our results show that the high energy momentum dependent continuum of
magnetic excitations observed in \LCO is a general feature of monolayer
cuprates.  A direct comparison between the magnetic excitation spectra measured
by INS and by RIXS reveals significant contributions from higher energy
excitations in the RIXS spectra that were not fully considered in the previous
analysis. This resulted in an overestimation of the zone-boundary energy at $(1/2,0)$.
As the energy resolution and theoretical understanding of the RIXS
cross-section continues to improve RIXS may become an ideal probe of this high
energy momentum dependent continuum of excitations.

\begin{acknowledgments}
We would like to thank John Hill, Mark Dean, and Jeroen van den Brink for
thoughtful comments and critical reading of this manuscript.  We are also
thankful to Michel Gingras for insightful discussions.  Work at the
University of Toronto was supported by NSERC of Canada. Work at Oak Ridge
National Laboratory's Spallation Neutron Source was sponsored by the Scientific
User Facilities Division, Office of Basic Energy Sciences, U.S.  Department of
Energy.  K.W.\@ Plumb acknowledges the support of the Ontario Graduate
Scholarship.  
\end{acknowledgments}

\end{document}